\documentclass[aps,prb,10pt, twocolumn, superscriptaddress, longbibliography, nobalancelastpage]{revtex4-1}
\UseRawInputEncoding
\usepackage{graphicx}
\usepackage{graphics}
\usepackage{amsmath}
\usepackage{amssymb}
\usepackage{amsfonts}
\usepackage{dsfont}
\usepackage{braket}
\usepackage{color}
\usepackage{braket,slashed}
\usepackage[mathscr]{euscript}
\usepackage{hyperref}
\usepackage{subfigure}
\usepackage{xfrac}
\usepackage{bm}
\usepackage{kantlipsum}
\usepackage{enumitem}
\usepackage{tikz}
\usepackage{geometry}
\usepackage{framed}
\usepackage{graphicx}
\usepackage{subfigure}

\allowdisplaybreaks[1]

\newcommand{\code}[1]{\texttt{#1}}

\usepackage{afterpage}
\usepackage{cleveref}

\newcommand{\be}{\begin{equation}}
\newcommand{\ee}{\end{equation}}
\newcommand{\bea}{\begin{eqnarray}}
\newcommand{\eea}{\end{eqnarray}}

\usepackage{xcolor}

\usepackage{lipsum}

\begin{document}
\title{Entanglement scaling for $\lambda\phi_2^4$}

\author{Bram Vanhecke}
\affiliation{Department of Physics and Astronomy, University of Ghent, Krijgslaan 281, 9000 Gent, Belgium}

\author{Frank Verstraete}
\affiliation{Department of Physics and Astronomy, University of Ghent, Krijgslaan 281, 9000 Gent, Belgium}

\author{Karel Van Acoleyen}
\affiliation{Department of Physics and Astronomy, University of Ghent, Krijgslaan 281, 9000 Gent, Belgium}

\begin{abstract}
	We study the $\lambda\phi^4$ model in $0+2$ dimensions at criticality, focusing on the scaling properties originating from the UV and IR physics. We demonstrate that the entanglement entropy, the correlation length $\xi$ and order parameters $\phi$ and $\phi^3$ exhibit distinctive double scaling properties that prove a powerful tool in the data analysis. The calculations are performed with boundary matrix product state methods on tensor network representations of the partition function to which the entanglement scaling hypothesis is applied, though the technique is equally applicable outside the realm of tensor networks. We find the value $\alpha_c=11.09698(31)$ for the critical point, improving on previous results.
\end{abstract}
\maketitle

\section{Introduction}

Scaling is one of the most profound concepts in modern day physics, as it plays a crucial role in the understanding and simulation of many-body systems that exhibit critical infrared (IR) behavior\cite{Fisher1972, Brezin1982, cardy1988}. Furthermore, for quantum field theories (QFTs) defined through lattice regularization, the continuum limit emerges precisely in the scaling behavior towards the ultraviolet (UV) critical point\cite{Luscher1991,Jansen1995,Wilson1973,Kogut1979}. In this context, QFTs with a second order phase transition --of which $D=2$ $\lambda \phi^4$ is the archetypal example-- hold a particular place. They are subject to both types of scaling, with the UV scaling defining the continuum limit, and the IR scaling near the QFT phase transition, each characterized by their own distinct CFT.

The typical procedure to study such a model with double critical behavior is to choose some values of the UV cut-off, for each of them determine the effective critical point, and extrapolate. Calculating all these critical points goes exactly as one would for any lattice model, typically using the IR scaling properties, through distinct power-laws or scaling hypotheses\cite{Loinaz1997,Korzec2011,Bronzin2018,Schaich2009}. It should thus be clear that each of those effective critical points are expensive to calculate, requiring many different values of the coupling and the IR cut-off (typically system size $L$, or in tensor network studies a bond dimension $\chi$). 
Further more, such an approach leaves questions about the interplay between the UV and IR CFTs untouched and all its possible benefits unused.\par
In this paper we will investigate how one could go about leveraging both the UV and IR scaling properties, to simultaneously use all data points in one fit for the continuum critical point. The technique builds and improves upon a previous work\cite{Vanhecke2019} and entails constructing quantities that are scale invariant w.r.t. both the IR and UV scaling, using them to effectuate a double collapse of all the data. This proves an effective way to fit the data and also captures and visualizes the ways that the UV and IR CFTs manifest themselves.\par
Numerical calculations are done within the tensor network (TN) framework: The regularized model is expressed as a square lattice TN that is contracted by determining the approximate matrix product state\cite{Cirac2020} (MPS) fixed point of the matrix product operator\cite{Haegeman2017} (MPO) transfer matrix, with the vumps\cite{fishman2018} algorithm. The finite bond dimension of the MPS introduces finite entanglement effects similar to finite size effects\cite{Nishino1996b, Pollmann2009, Tagliacozzo2008,pirvu2012matrix}. In the entanglement scaling hypothesis\cite{Vanhecke2019} for MPS, a quantity, $\delta$, is identified which acts as an inverse system size $L^{-1}$ under scaling, and can be used as a substitute for $\chi$ to label the MPS results. The finite bond dimension effects will thus be handled using the scaling properties, in exactly the same way as one would for calculations performed at finite size\cite{cardy1988}. Our analysis can thus be straightforwardly applied to Hamiltonian methods\cite{milsted2013} or approaches based on the corner transfer matrix method (CTM) \cite{Nishino1996,Orus2009,Corboz2011}, for which the entanglement scaling hypothesis also holds, and furthermore our method can be trivially adapted to methods with finite lattice size effects like Monte Carlo or exact diagonalization.\par
We will first review the model, then construct the aforementioned scale invariant quantities, and finally, we discuss the results obtained by optimizing the collapses.

\section{The Model}
We start from the euclidean action:
\begin{equation}
	\mathcal{L}(\phi)=\frac{1}{2}\partial_\mu\phi\partial^\mu\phi+\frac{1}{2}\mu_0^2\phi^2+\frac{1}{4}\lambda_0\phi^4
\end{equation}
where $\phi$ is a real function of $2$-$D$ space. This is a superrenormalizable QFT at the perturbative level\cite{Chang1976} and it has been proven\cite{Brydges1983} that this model gives rise to a non-trivial theory at the full non-perturbative level. The model has a $\mathbb{Z}_2$ symmetry breaking phase transition in the Ising universality class, at a coupling that is beyond the reach of standard perturbation theory.\par
We study this model using lattice regularization, discretizing both space and time. 
\begin{equation}
	Z=\int\prod_{i}\text{d}\phi_ie^{-\sum\limits_{\braket{i,j}}\frac{1}{2}(\phi_i-\phi_j)^2-\sum\limits_i\frac{1}{2}\mu^2\phi_i^2+\frac{1}{4}\lambda\phi_i^4}\,,
	\label{eq:partitionfunction}
\end{equation}
The above partition function can be written as a tensor network of finite bond dimension by discretizing $\phi_i$, the details of which are in the supplemental material. Different than previous approaches \cite{delcamp2020,Kadoh2018}, our approach is distinguished by arbitrary precision, optimal bond dimension, and minimal computational cost.\par

\section{Continuum Limit}
\begin{figure}
	\centering
	\includegraphics[width=0.75\linewidth]{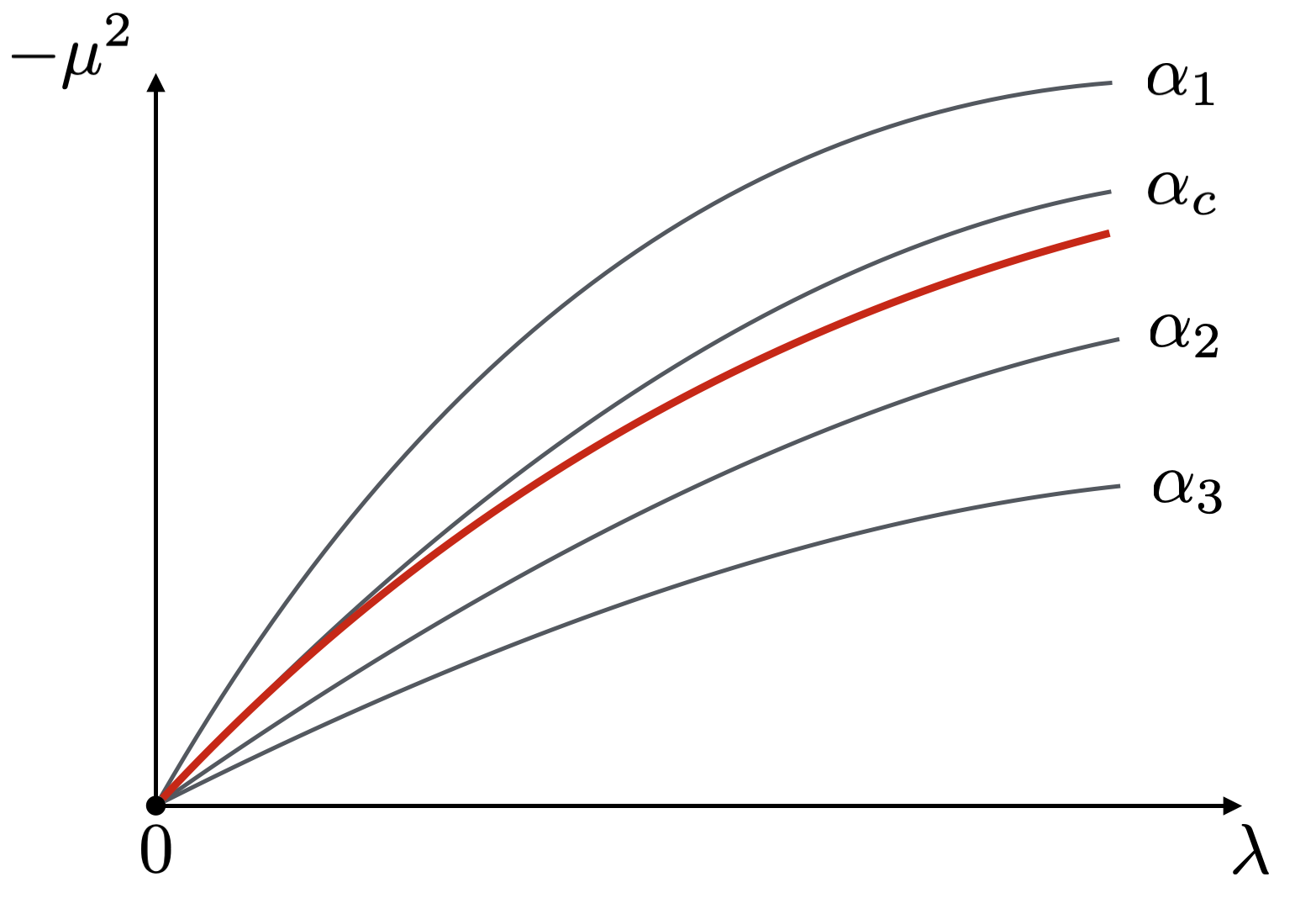}
	\caption{Sketch of the phase diagram of the lattice model (\ref{eq:partitionfunction}) in the $(-\mu^2,\lambda)$ plane, with the critical line in red, and also some lines describing different continuum limits (eq. \ref{eq:continuumlimit}). The prime goal of our simulations is to find the critical $\alpha_c$,  for which the continuum limit converges to the critical line for $a\rightarrow 0$. }
	\label{fig:continuumlimit}
\end{figure}

To extract the continuum theory from this lattice model one should vary the parameters $\mu$ and $\lambda$ such that every conceivable linear length scale (i.e. masses, scattering lengths,...) becomes proportional to all others as they diverge. From perturbation theory we get the precise prescription for taking this QFT continuum limit \cite{Kadoh2018}:\begin{equation}
	\begin{split}
		\mu^2=&\lambda \alpha -3\lambda A(\lambda \alpha)\\
		&A(x)=\int_0^\pi \int_0^\pi  \frac{ \text{d}y \text{d}z}{x+4\sin(y)^2+4\sin(z)^2}\\
		a^2=&\lambda
	\end{split}
	\label{eq:continuumlimit}
\end{equation}
where $a$ is the effective lattice spacing in real space units. The function $A(x)$ originates from the (one-loop) tadpole diagram of the mass renormalization for our particular lattice regularization. For each $\alpha$ the above expression parameterizes a continuum limit, prescribing how $\mu$ and $\lambda$ should be varied to take the effective lattice spacing $a$ to zero while preserving all correlations in real space units. See the illustration in Fig. \ref{fig:continuumlimit}. It is important to realize that the parameter $\alpha$ is universal, in the sense that it may be compared across different regularization schemes, it is usually denoted $\mu_R/\lambda$.\par

The presence of a second order phase transition of the QFT implies that the continuum correlation length $=\lim_{a\to 0}a\xi(a,\alpha)$ diverges as $\alpha$ is tuned to the critical value $\alpha_c$. For the lattice model of Eq. \ref{eq:partitionfunction} this means there must be a critical line in the $(\mu^2,\lambda)$-plane that is parameterized by $\alpha_c$, up to leading order in $a$. However, at $a\gg0$, there will be finite cut-off effects that make this critical line deviate from a curve of constant $\alpha$, effectively causing the critical value of $\alpha$ to shift with $\lambda$. This is illustrated in Fig. \ref{fig:continuumlimit}. 
\par 
This issue cannot be overcome by simply working with small enough $\lambda$ as the gains in decreased finite cut-off effects are vastly outweighed by the added computational cost
. We will thus work with small but reasonable $\lambda$ data and fit the subleading corrections that should be added to the definition of $\alpha$, in Eq. \ref{eq:continuumlimit}, that are required to make the critical point constant with $\lambda$. \par 
First, though, we will ignore these complications and build a scaling theory around $a\approx0$ and $\alpha\approx\alpha_c$, and later add the necessary corrections. We will use the scaling properties of the UV, which follow from the existence of a continuum limit, and the scaling properties of the IR, caused by the second order phase transition, to collapse 3-D data $O(\mu^2,\lambda,\chi)$ to a 1-D curve $\tilde{O}(\alpha)$. \par

\section{Double scaling: UV and IR}
To do a UV scale transformation one should imagine doing an RG transformation as generally appearing in QFT: rescaling the cut-off, here the lattice spacing $a$, while keeping the continuum quantities fixed. The exponents of all the variables and observables are then determined by their response to such a transformation (up to leading order in $a$).\par
It thus follows from Eq. \ref{eq:continuumlimit} that $\lambda$ has UV scaling exponent $2$, and $\alpha$ has exponent $0$
. The lattice-correlation length $\xi$ has UV exponent $-1$, since $a\xi$ --the continuum correlation length-- must remain constant when varying $a$. And, similarly the inverse linear system size $L^{-1}$ (in lattice units), or its MPS counterpart $\delta$, has exponent $1$.\par

Since there is no wave-function renormalization needed for $\phi_2^4$ in the continuum limit, the UV exponent of the field $\phi$ is simply $0$, corresponding to its canonical dimension. To extend the number of observables, we have also considered the composite operator $\phi^3$. As such this operator is not properly defined for the QFT, as it has a UV divergence, arising from the tadpole contribution to the disconnected part, which evaluates to $3\braket{\phi^2}\times\phi\sim -\frac{3}{4\pi}\log(\lambda)\phi$ (see supplemental material). We therefore subtract this divergence to define a regularized $\tilde{\phi}^3=\phi^3+\frac{3}{4\pi}\log(\lambda)\phi$. This finite operator $\tilde{\phi}^3$ then scales according to its canonical dimension, which is again 0.\par 
Finally, we have also considered the entanglement entropy $S$ as a QFT 'observable'\footnote{Here we use the term 'observables' for quantities exhibiting a well posed continuum limit. The entanglement entropy is not an observable in the quantum information theoretic sense \cite{Nielsen-Chuang} }. From the diverging lattice correlation length at fixed $\alpha$ in the $a\to 0$ limit, and the Cardy-Calabrese\cite{Calabrese_2004} entanglement law we can anticipate $S\sim-\frac{c_{uv}}{6}\log(a)$, with $c_{uv}$ the central charge of the free boson CFT $c_{uv}=1$. By formally considering the quantity $e^S$, we can translate this logarithmic scaling to a UV-exponent $-\frac{c_{uv}}{6}$. 

\begin{table}
	\begin{center}
		\begin{tabular}{| c | c | c |}	
			\hline
			& UV exponent & IR exponent \\ \hline
			$\lambda$ & 2 & 0 \\ \hline
			$\alpha-\alpha_c$ & 0 & $1/\nu=1$ \\ \hline
			$L^{-1}\,,\,\delta$  & 1 & 1 \\ \hline
			$\xi$  & -1 & -1 \\ \hline
			$\phi$  & 0 & $\beta=1/8$ \\ \hline
			$\phi^3-\frac{3}{4\pi}\log(\lambda)\phi$  & 0 & $\beta=1/8$ \\ \hline
			$\exp(S)$& $-\frac{c_{uv}}{6}=-\frac{1}{6}$ & $-\frac{c_{ir}}{6}=-\frac{1}{12}$ \\ \hline
		\end{tabular}
	\end{center}
	\caption{Summary of all the UV and IR scaling exponents}
	\label{table:dimensions}
\end{table}

Next, we consider the IR scaling, which can be understood as an RG-flow of the continuum theory. The IR exponent of $\lambda$ is $0$ as the continuum theory is independent of the lattice spacing $a=\sqrt{\lambda}$. The exponent $1/\nu$ of $\alpha-\alpha_c$ acts as temperature does in the Ising model
. Regular lengths have their usual exponent, so the correlation length has IR exponent $-1$, and $L^{-1}$ and $\delta$ have exponent $1$. As was found for the UV scaling of $e^S$, we similarly find that the IR exponent of $e^S$ should be $-\frac{c_{ir}}{6}$, where $c_{ir}=1/2$ is the central charge of the Ising CFT. Finally, 
the observables $\phi$ and $\phi^3$, and thus also $\tilde{\phi}^3$, act as $\mathbb{Z}_2$ order parameters with respect to the Ising critical point. They therefore have IR exponent $\beta$, which in this case is $1/8$.
See Table \ref{table:dimensions}, for all UV and IR exponents.\par

We are now ready to construct scale invariant quantities from our 4 observables $O=\xi, \phi,\tilde{\phi}^3, e^{S}$. This is achieved by compensating for the UV exponent with appropriate factors of $\sqrt{\lambda}$ (the lattice spacing) and similarly for the IR exponent with factors of $\Delta=\delta/\sqrt{\lambda}$ (acting as an inverse system size in physical units) to construct an IR and UV scale invariant object $\mathcal{O}$:
\begin{equation}
	\mathcal{O} = \lambda^{-d_{uv}/2}\Delta^{-d_{ir}}O\,.
	\label{eq:Onocorrections}
\end{equation}\par 
Similarly we construct from $\alpha-\alpha_c$ a scale invariant quantity: $\Delta^{-1/\nu}(\alpha-\alpha_c)$.\par 
A general data point consists of variables: $\mu^2$, $\lambda$, and the bond dimension $\chi$, that map to an observable $O$. We then do a change of variable of $\chi$ to $\delta$ or $\Delta=\delta/\sqrt{\lambda}$ with improved scaling properties\cite{Vanhecke2019}. Next, those four numbers are transformed to $\Delta^{-1/\nu}(\alpha-\alpha_c)$, $\lambda$, and $\alpha-\alpha_c$, that map to $\mathcal{O}$, representing a function that, by construction, does not have an explicit dependence on $\lambda$, or $\alpha-\alpha_c$. If everything works out, the 4-D data can hence be collapsed to a 2-D curve:
\begin{equation}
	[\mu^2,\,\lambda,\,\chi,\,O]\to[\Delta^{-1/\nu}(\alpha-\alpha_c),\,\lambda^{-d_{uv}/2}\Delta^{-d_{ir}}O]\,.
	\label{eq:doublecollapse}
\end{equation}

\section{Corrections}
The critical point depends strongly on $\lambda$ when simply using the definition in Eq.\ref{eq:continuumlimit}, which is problematic for the UV- \emph{and} IR-scaling that we hope to impose. This can be solved by adding corrections to $\alpha$ that make the lattice critical point $\alpha_c$ approximately constant as a function of $\lambda$, so that for the newly defined $\alpha$ the critical (red) line and $\alpha_c$ line inf Fig. \ref{fig:continuumlimit} become identical.\par
We provide $\alpha$ with $\lambda$-corrections parameterized as follows:
\begin{equation}
	\begin{split}
		\alpha\to&\alpha+A\lambda\log(\lambda)+B\lambda\\
		&+C\lambda^2\log(\lambda)^2+D\lambda^2\log(\lambda)+E\lambda^2+\dots\,.
		\label{eq:alphacorr}
	\end{split}
\end{equation}
The first two terms have been observed before, see e.g.\cite{Schaich2009,delcamp2020}, and in the supplemental material we show that the mass renormalization from the two-loop setting sun diagram with the lattice regularization (\ref{eq:partitionfunction}) indeed produces terms of the form $\lambda\log\lambda/\lambda_0$. The subsequent terms in the series are simply products of the first two. \par
We also expect corrections for the observables $O$, since the above described scaling properties only hold true in the limit $\lambda \to 0$ and $\alpha-\alpha_c\to 0$. We give the same type of corrections as we considered for $\alpha$:
\begin{equation}
	\begin{split}
		O\to&O\,\Big(A_O+B_O\lambda\log(\lambda)+C_O\lambda\\
		&+D_O\lambda^2\log(\lambda)^2+E_O\lambda^2\log(\lambda)+F_O\lambda^2+\dots\Big)\,.
		\label{eq:Ocorr}
	\end{split}
\end{equation}
\par
These prefactors $A_O,B_O,\dots$ can be given an $\alpha-\alpha_c$ dependence, to provide corrections to the IR behavior. In practice, it will only be needed to give $A_O$ a linear dependence on $\alpha-\alpha_c$, and then only for the entanglement entropy. This type of correction to the IR scaling can be understood as compensating for the following generic form of the power law:
\begin{equation}
	O\sim (\alpha-\alpha_c)^{\beta/\nu}\Big(1+A(\alpha-\alpha_c)^\omega+\dots\Big)\,.
\end{equation}
\par 
The point is now that we have to adjust the parameters $A,B,\cdots$ of Eq. \ref{eq:alphacorr} and Eq. \ref{eq:Ocorr} to make the scaling properties in Table \ref{table:dimensions} hold true in the entire data set; in other words, they will be optimized such as to effectuate an optimal collapse of the data.
\begin{figure}
	\centering
	\includegraphics[width=0.85\linewidth]{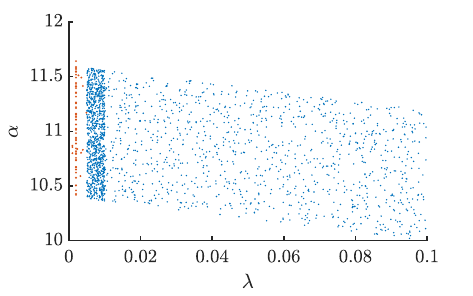}
	\caption{A plot of the $\lambda$ and $\alpha$ values used in the fits. The red bar indicates the part of the data that is not included in the preliminary fits and is used to estimate the out-of-sample error.}
	\label{fig:dataset}
\end{figure}
\begin{figure}
	\centering
	\includegraphics[width=0.85\linewidth]{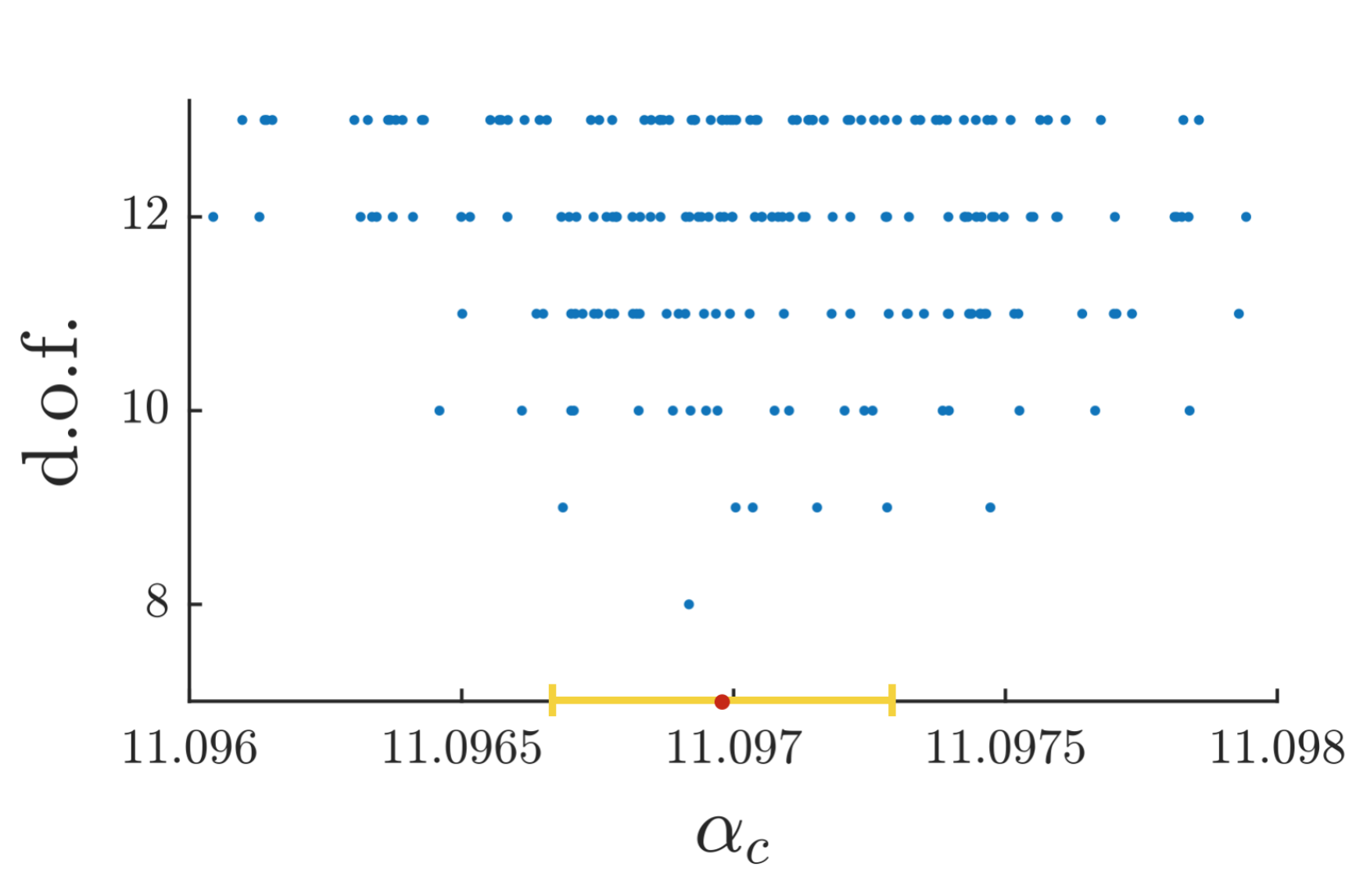}
	\caption{For the 268 'good' fits (see text) : the value of $\alpha_c$ plotted versus the number of fit parameters that were included ($\alpha_c + $ the parameters in the scaling corrections). We use the median to estimate the best value of $\alpha_c=11.09698(31)$ (indicated with a red dot), and an error bar is estimated with the median distance from that best value (yellow interval).}
	\label{fig:alphacplot}
\end{figure}

\begin{figure*}
	\centering
	\includegraphics[width=0.85\linewidth]{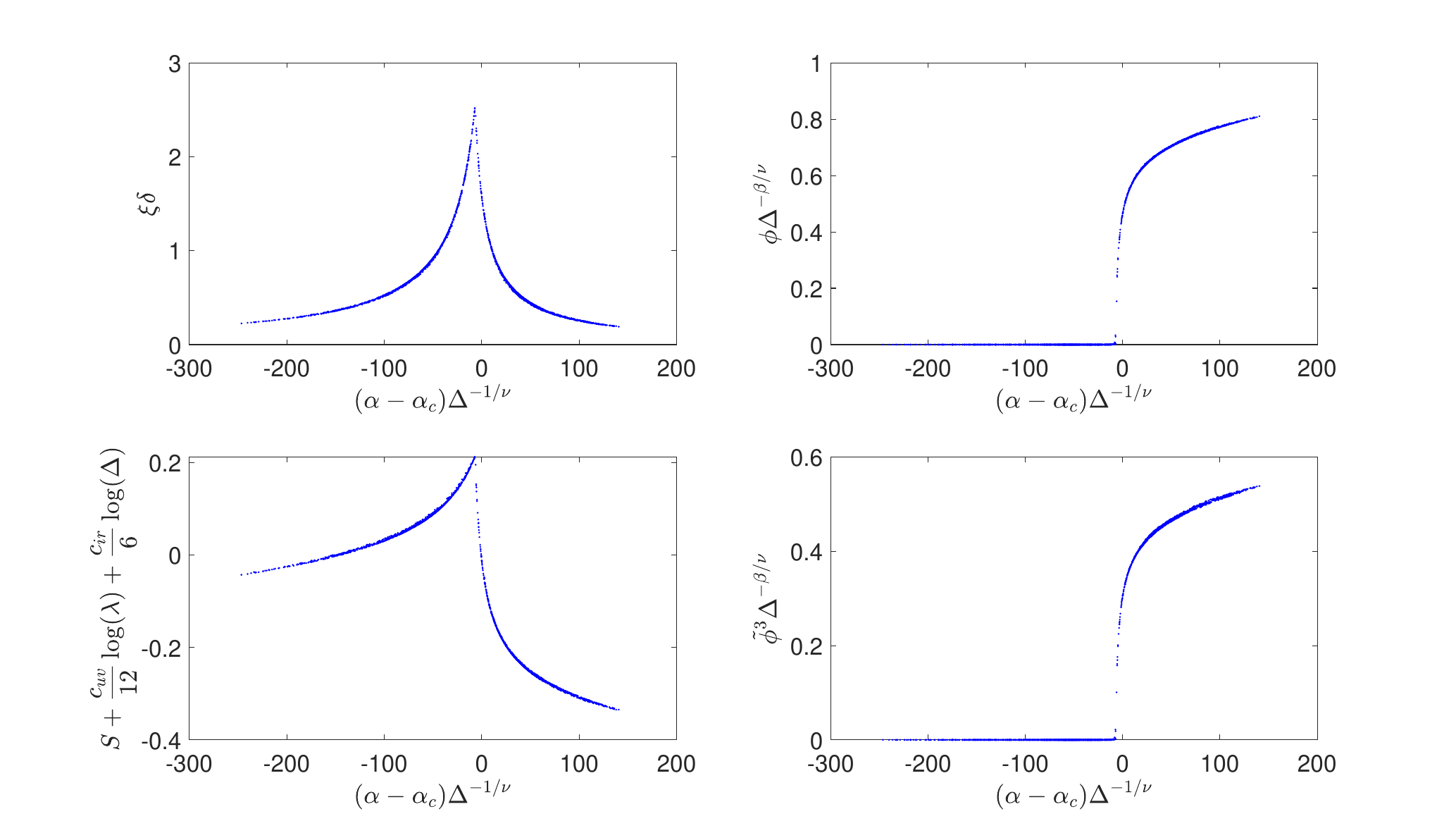}
	\caption{TOP-LEFT: The rescaled correlation length $\xi$ vs the rescaled coupling $\alpha-\alpha_c$. TOP-RIGHT: The order parameter $\phi$ vs the rescaled coupling $\alpha-\alpha_c$.BOTTOM-LEFT: The renormalized and rescaled order parameter $\tilde{\phi}^3$ vs the rescaled coupling $\alpha-\alpha_c$.BOTTOM-RIGHT: The logarithmically rescaled entanglement entropy $S$ vs the rescaled coupling $\alpha-\alpha_c$. Note that the shifted singular points of these functions imply a shift in location of the effective critical point by an amount proportional to $\Delta^{-1/\nu}$, this is a general feature found in scaling functions\cite{Vanhecke2019}}
	\label{fig:fit}
\end{figure*}

\section{Fitting Procedure}
We plot the four rescaled observables $\xi$, $\phi$, $\tilde{\phi}^3$ and $S$ versus the rescaled parameter $\Delta^{-1/\nu}(\alpha-\alpha_c)$ (see eq. \ref{eq:doublecollapse}), and optimize the average orthogonal distance from those data points to a fit function. The practical details of this are straightforward and presented in the supplemental material.
\par 
Our goal is to optimize the corrections (eqs. (\ref{eq:alphacorr}) and \ref{eq:Ocorr})) discussed in the previous section, and use the $\lambda>0.001$ data to extrapolate to $\lambda=0$. There is, however, a clear danger of including too many corrections that will over-optimize the fit for $\lambda\in[0.001,0.1]$, leading to a bad extrapolation $\lambda\to 0$. Conversely, the same thing can happen if not enough corrections are included.\par 
To remedy this, we perform all our fits, with various combinations of corrections included, using all but the smallest $\lambda$ data, indicated in Fig. \ref{fig:dataset}, allowing us to see which fits permit extrapolation to smaller $\lambda$. Specifically, our criterion for a 'good' set of corrections is that they give a smaller average out-of-sample error than in-sample, for all observables simultaneously. 

\section{Results}
We calculated 2081 data points, each with a random $\lambda\in[0.001,0.1]$, associated random $\alpha$ chosen close to the expected critical point, and a random MPS bond dimension $\chi\in[100, 200]$. 
\par 
We considered $372$ different combinations of corrections (\ref{eq:alphacorr}) and (\ref{eq:Ocorr}), deemed reasonable from preliminary fits but varied enough to be unbiased to any specific set of corrections. From this we found $268$ fits with a smaller out-of-sample than in-sample error. For all these 'good' fits we optimize again, this time using all the data, and plot the values for $\alpha_c$ versus the number of fitting parameters in Fig. \ref{fig:alphacplot}.\par

For a single fit, there is no clear way to estimate the error on $\alpha_c$ with this technique. However, we can use the fact that there are many different fits to get an idea of the error bar, but notice that these different results are not strictly statistical independent. For our final value of $\alpha_c$ we take the median value of all good fits, while for the quoted error bar we take the median distance from this median value of $\alpha_c$.  This final result is compared with previous results in Table. \ref{table:alphacfits} and the collapse plots for the best fit is shown in Fig.\ref{fig:fit}, and the specific parameters used may be found in the supplemental material.\par

\par 
From this set of 'good' fits, we can conclude that we absolutely need to include $\lambda\log\lambda$ and $\lambda^2\log\lambda$ terms (those terms were not included in the previous work\cite{Vanhecke2019} and are responsible for the sub-optimal results there), but it also became clear that the $\lambda^2\log(\lambda)^2$,  $\lambda^3\log(\lambda)^3$ and $\lambda^3\log(\lambda)^2$ terms should not be included. $\lambda^3\log(\lambda)$ and $\lambda^3$ could help. but could not improve the out-of-sample error w.r.t. the set of corrections shown above. 
\begin{table}
	\begin{center}
		\begin{tabular}{| c | c | c |}	
			\hline
			Method & Year & $\alpha_c $ \\ \hline
			MPS\cite{milsted2013} & 2013 &  11.064(20) \\ \hline
			Hamiltonian truncation\cite{Elias-Miro:2017} & 2017 &  11.04(12) \\ \hline
			Borel resummation\cite{serone2018} & 2018 &   11.23(14) \\ \hline
			Monte Carlo\cite{Bronzin2018} & 2018 &  11.055(20) \\ \hline
			TRG\cite{Kadoh2018} & 2019 & 10.913(56) \\ \hline
			gilt-TNR\cite{delcamp2020} & 2020 &  11.0861(90) \\ \hline
			This work & 2021 & 11.09698(31) \\ \hline
		\end{tabular}
	\end{center}
	\caption{Comparison with some results from the literature.}
	\label{table:alphacfits}
\end{table}

It is interesting to check whether the fit would allow for the calculation of the prefactors of the universal divergent terms (as usually determined by Feynman diagrams). If we fit the renormalization of the $\phi^3$, we find a correction $-0.23888(65)\log(\lambda)\phi$, which should be compared with the analytical $\frac{3}{4\pi}\approx0.23873$ used previously.

\section{Conclusion}
A critical QFT is something special. Like with any QFT the regulated model becomes UV critical as the cut-off is taken to infinity, but there is also an IR criticality due to the continuum phase transition. It thus has a sort of double criticality, each with a different CFT description. In this work we have shown how, in the case of $\lambda \phi^4_2$, this feature allows for a double scaling analysis, giving rise to collapse plots for the different QFT observables, encapsulating both scaling behaviors in one go. We stress that our approach is general, in the sense that it can be applied to any method that is confronted with both types of scaling, and is not limited to tensor network based methods. Furthermore, the double scaling approach is not restricted to $D=2$ space-time dimensions or zero temperature. In particular it would be interesting to explore the double scaling also for $D>2$ QFTs that exhibit a purported continuous phase transition, e.g. for Monte-Carlo simulations of the finite temperature chiral symmetry breaking transition for $N_f=2$ massless QCD.       


\section{Acknowledgments}
We would like to thank Rui-Zhen Huang for an interesting discussion on scale invariant quantities. This research is supported by ERC grant QUTE (647905) (BV, FV)


\begin{thebibliography}{33}%
	\makeatletter
	\providecommand \@ifxundefined [1]{%
		\@ifx{#1\undefined}
	}%
	\providecommand \@ifnum [1]{%
		\ifnum #1\expandafter \@firstoftwo
		\else \expandafter \@secondoftwo
		\fi
	}%
	\providecommand \@ifx [1]{%
		\ifx #1\expandafter \@firstoftwo
		\else \expandafter \@secondoftwo
		\fi
	}%
	\providecommand \natexlab [1]{#1}%
	\providecommand \enquote  [1]{``#1''}%
	\providecommand \bibnamefont  [1]{#1}%
	\providecommand \bibfnamefont [1]{#1}%
	\providecommand \citenamefont [1]{#1}%
	\providecommand \href@noop [0]{\@secondoftwo}%
	\providecommand \href [0]{\begingroup \@sanitize@url \@href}%
	\providecommand \@href[1]{\@@startlink{#1}\@@href}%
	\providecommand \@@href[1]{\endgroup#1\@@endlink}%
	\providecommand \@sanitize@url [0]{\catcode `\\12\catcode `\$12\catcode
		`\&12\catcode `\#12\catcode `\^12\catcode `\_12\catcode `\%12\relax}%
	\providecommand \@@startlink[1]{}%
	\providecommand \@@endlink[0]{}%
	\providecommand \url  [0]{\begingroup\@sanitize@url \@url }%
	\providecommand \@url [1]{\endgroup\@href {#1}{\urlprefix }}%
	\providecommand \urlprefix  [0]{URL }%
	\providecommand \Eprint [0]{\href }%
	\providecommand \doibase [0]{http://dx.doi.org/}%
	\providecommand \selectlanguage [0]{\@gobble}%
	\providecommand \bibinfo  [0]{\@secondoftwo}%
	\providecommand \bibfield  [0]{\@secondoftwo}%
	\providecommand \translation [1]{[#1]}%
	\providecommand \BibitemOpen [0]{}%
	\providecommand \bibitemStop [0]{}%
	\providecommand \bibitemNoStop [0]{.\EOS\space}%
	\providecommand \EOS [0]{\spacefactor3000\relax}%
	\providecommand \BibitemShut  [1]{\csname bibitem#1\endcsname}%
	\let\auto@bib@innerbib\@empty
	\bibitem [{\citenamefont {Fisher}\ and\ \citenamefont
		{Barber}(1972)}]{Fisher1972}%
	\BibitemOpen
	\bibfield  {author} {\bibinfo {author} {\bibfnamefont {Michael~E.}\
			\bibnamefont {Fisher}}\ and\ \bibinfo {author} {\bibfnamefont {Michael~N.}\
			\bibnamefont {Barber}},\ }\bibfield  {title} {\enquote {\bibinfo {title}
			{Scaling theory for finite-size effects in the critical region},}\ }\href
	{\doibase 10.1103/PhysRevLett.28.1516} {\bibfield  {journal} {\bibinfo
			{journal} {Phys. Rev. Lett.}\ }\textbf {\bibinfo {volume} {28}},\ \bibinfo
		{pages} {1516--1519} (\bibinfo {year} {1972})}\BibitemShut {NoStop}%
	\bibitem [{\citenamefont {Br\'{e}zin}(1982)}]{Brezin1982}%
	\BibitemOpen
	\bibfield  {author} {\bibinfo {author} {\bibfnamefont {E.}~\bibnamefont
			{Br\'{e}zin}},\ }\bibfield  {title} {\enquote {\bibinfo {title} {An
				investigation of finite size scaling},}\ }\href {\doibase
		10.1051/jphys:0198200430101500} {\bibfield  {journal} {\bibinfo  {journal}
			{J. Phys. France}\ }\textbf {\bibinfo {volume} {43}},\ \bibinfo {pages} {15}
		(\bibinfo {year} {1982})}\BibitemShut {NoStop}%
	\bibitem [{\citenamefont {Cardy}(1988)}]{cardy1988}%
	\BibitemOpen
	\bibfield  {author} {\bibinfo {author} {\bibfnamefont {J.L.}\ \bibnamefont
			{Cardy}},\ }\href {https://books.google.be/books?id=lqPvAAAAMAAJ} {\emph
		{\bibinfo {title} {Finite-size Scaling}}},\ Current physics\ (\bibinfo
	{publisher} {North-Holland},\ \bibinfo {year} {1988})\BibitemShut {NoStop}%
	\bibitem [{\citenamefont {Luscher}\ \emph {et~al.}(1991)\citenamefont
		{Luscher}, \citenamefont {Weisz},\ and\ \citenamefont {Wolff}}]{Luscher1991}%
	\BibitemOpen
	\bibfield  {author} {\bibinfo {author} {\bibfnamefont {Martin}\ \bibnamefont
			{Luscher}}, \bibinfo {author} {\bibfnamefont {Peter}\ \bibnamefont {Weisz}},
		\ and\ \bibinfo {author} {\bibfnamefont {Ulli}\ \bibnamefont {Wolff}},\
	}\bibfield  {title} {\enquote {\bibinfo {title} {{A Numerical method to
					compute the running coupling in asymptotically free theories}},}\ }\href
	{\doibase 10.1016/0550-3213(91)90298-C} {\bibfield  {journal} {\bibinfo
			{journal} {Nucl. Phys.}\ }\textbf {\bibinfo {volume} {B359}},\ \bibinfo
		{pages} {221--243} (\bibinfo {year} {1991})}\BibitemShut {NoStop}%
	\bibitem [{\citenamefont {Jansen}\ \emph {et~al.}(1996)\citenamefont {Jansen},
		\citenamefont {Liu}, \citenamefont {Luscher}, \citenamefont {Simma},
		\citenamefont {Sint}, \citenamefont {Sommer}, \citenamefont {Weisz},\ and\
		\citenamefont {Wolff}}]{Jansen1995}%
	\BibitemOpen
	\bibfield  {author} {\bibinfo {author} {\bibfnamefont {Karl}\ \bibnamefont
			{Jansen}}, \bibinfo {author} {\bibfnamefont {Chuan}\ \bibnamefont {Liu}},
		\bibinfo {author} {\bibfnamefont {Martin}\ \bibnamefont {Luscher}}, \bibinfo
		{author} {\bibfnamefont {Hubert}\ \bibnamefont {Simma}}, \bibinfo {author}
		{\bibfnamefont {Stefan}\ \bibnamefont {Sint}}, \bibinfo {author}
		{\bibfnamefont {Rainer}\ \bibnamefont {Sommer}}, \bibinfo {author}
		{\bibfnamefont {Peter}\ \bibnamefont {Weisz}}, \ and\ \bibinfo {author}
		{\bibfnamefont {Ulli}\ \bibnamefont {Wolff}},\ }\bibfield  {title} {\enquote
		{\bibinfo {title} {{Nonperturbative renormalization of lattice QCD at all
					scales}},}\ }\href {\doibase 10.1016/0370-2693(96)00075-5} {\bibfield
		{journal} {\bibinfo  {journal} {Phys. Lett.}\ }\textbf {\bibinfo {volume}
			{B372}},\ \bibinfo {pages} {275--282} (\bibinfo {year} {1996})}\BibitemShut
	{NoStop}%
	\bibitem [{\citenamefont {Wilson}\ and\ \citenamefont
		{Kogut}(1974)}]{Wilson1973}%
	\BibitemOpen
	\bibfield  {author} {\bibinfo {author} {\bibfnamefont {K.~G.}\ \bibnamefont
			{Wilson}}\ and\ \bibinfo {author} {\bibfnamefont {John~B.}\ \bibnamefont
			{Kogut}},\ }\bibfield  {title} {\enquote {\bibinfo {title} {{The
					Renormalization group and the epsilon expansion}},}\ }\href {\doibase
		10.1016/0370-1573(74)90023-4} {\bibfield  {journal} {\bibinfo  {journal}
			{Phys. Rept.}\ }\textbf {\bibinfo {volume} {12}},\ \bibinfo {pages} {75--199}
		(\bibinfo {year} {1974})}\BibitemShut {NoStop}%
	\bibitem [{\citenamefont {Kogut}(1979)}]{Kogut1979}%
	\BibitemOpen
	\bibfield  {author} {\bibinfo {author} {\bibfnamefont {John~B.}\ \bibnamefont
			{Kogut}},\ }\bibfield  {title} {\enquote {\bibinfo {title} {An introduction
				to lattice gauge theory and spin systems},}\ }\href {\doibase
		10.1103/RevModPhys.51.659} {\bibfield  {journal} {\bibinfo  {journal} {Rev.
				Mod. Phys.}\ }\textbf {\bibinfo {volume} {51}},\ \bibinfo {pages} {659--713}
		(\bibinfo {year} {1979})}\BibitemShut {NoStop}%
	\bibitem [{\citenamefont {Loinaz}\ and\ \citenamefont
		{Willey}(1998)}]{Loinaz1997}%
	\BibitemOpen
	\bibfield  {author} {\bibinfo {author} {\bibfnamefont {Will}\ \bibnamefont
			{Loinaz}}\ and\ \bibinfo {author} {\bibfnamefont {R.~S.}\ \bibnamefont
			{Willey}},\ }\bibfield  {title} {\enquote {\bibinfo {title} {{Monte Carlo
					simulation calculation of critical coupling constant for continuum phi**4 in
					two-dimensions}},}\ }\href {\doibase 10.1103/PhysRevD.58.076003} {\bibfield
		{journal} {\bibinfo  {journal} {Phys. Rev. D}\ }\textbf {\bibinfo {volume}
			{58}},\ \bibinfo {pages} {076003} (\bibinfo {year} {1998})},\ \Eprint
	{http://arxiv.org/abs/hep-lat/9712008} {arXiv:hep-lat/9712008} \BibitemShut
	{NoStop}%
	\bibitem [{\citenamefont {Korzec}\ \emph {et~al.}(2011)\citenamefont {Korzec},
		\citenamefont {Vierhaus},\ and\ \citenamefont {Wolff}}]{Korzec2011}%
	\BibitemOpen
	\bibfield  {author} {\bibinfo {author} {\bibfnamefont {Tomasz}\ \bibnamefont
			{Korzec}}, \bibinfo {author} {\bibfnamefont {Ingmar}\ \bibnamefont
			{Vierhaus}}, \ and\ \bibinfo {author} {\bibfnamefont {Ulli}\ \bibnamefont
			{Wolff}},\ }\bibfield  {title} {\enquote {\bibinfo {title} {Performance of a
				worm algorithm in theory at finite quartic coupling},}\ }\href {\doibase
		10.1016/j.cpc.2011.03.018} {\bibfield  {journal} {\bibinfo  {journal}
			{Computer Physics Communications}\ }\textbf {\bibinfo {volume} {182}},\
		\bibinfo {pages} {1477--1480} (\bibinfo {year} {2011})}\BibitemShut {NoStop}%
	\bibitem [{\citenamefont {Bronzin}\ \emph {et~al.}(2019)\citenamefont
		{Bronzin}, \citenamefont {De~Palma},\ and\ \citenamefont
		{Guagnelli}}]{Bronzin2018}%
	\BibitemOpen
	\bibfield  {author} {\bibinfo {author} {\bibfnamefont {Simone}\ \bibnamefont
			{Bronzin}}, \bibinfo {author} {\bibfnamefont {Barbara}\ \bibnamefont
			{De~Palma}}, \ and\ \bibinfo {author} {\bibfnamefont {Marco}\ \bibnamefont
			{Guagnelli}},\ }\bibfield  {title} {\enquote {\bibinfo {title} {{New Monte
					Carlo determination of the critical coupling in \ensuremath{\phi}24
					theory}},}\ }\href {\doibase 10.1103/PhysRevD.99.034508} {\bibfield
		{journal} {\bibinfo  {journal} {Phys. Rev. D}\ }\textbf {\bibinfo {volume}
			{99}},\ \bibinfo {pages} {034508} (\bibinfo {year} {2019})},\ \Eprint
	{http://arxiv.org/abs/1807.03381} {arXiv:1807.03381 [hep-lat]} \BibitemShut
	{NoStop}%
	\bibitem [{\citenamefont {Schaich}\ and\ \citenamefont
		{Loinaz}(2009)}]{Schaich2009}%
	\BibitemOpen
	\bibfield  {author} {\bibinfo {author} {\bibfnamefont {David}\ \bibnamefont
			{Schaich}}\ and\ \bibinfo {author} {\bibfnamefont {Will}\ \bibnamefont
			{Loinaz}},\ }\bibfield  {title} {\enquote {\bibinfo {title} {{An Improved
					lattice measurement of the critical coupling in phi(2)**4 theory}},}\ }\href
	{\doibase 10.1103/PhysRevD.79.056008} {\bibfield  {journal} {\bibinfo
			{journal} {Phys. Rev. D}\ }\textbf {\bibinfo {volume} {79}},\ \bibinfo
		{pages} {056008} (\bibinfo {year} {2009})},\ \Eprint
	{http://arxiv.org/abs/0902.0045} {arXiv:0902.0045 [hep-lat]} \BibitemShut
	{NoStop}%
	\bibitem [{\citenamefont {{Vanhecke}}\ \emph {et~al.}(2019)\citenamefont
		{{Vanhecke}}, \citenamefont {{Haegeman}}, \citenamefont {{Van Acoleyen}},
		\citenamefont {{Vanderstraeten}},\ and\ \citenamefont
		{{Verstraete}}}]{Vanhecke2019}%
	\BibitemOpen
	\bibfield  {author} {\bibinfo {author} {\bibfnamefont {B.}~\bibnamefont
			{{Vanhecke}}}, \bibinfo {author} {\bibfnamefont {J.}~\bibnamefont
			{{Haegeman}}}, \bibinfo {author} {\bibfnamefont {K.}~\bibnamefont {{Van
					Acoleyen}}}, \bibinfo {author} {\bibfnamefont {L.}~\bibnamefont
			{{Vanderstraeten}}}, \ and\ \bibinfo {author} {\bibfnamefont
			{F.}~\bibnamefont {{Verstraete}}},\ }\bibfield  {title} {\enquote {\bibinfo
			{title} {Scaling hypothesis for matrix product states},}\ }\href {\doibase
		10.1103/PhysRevLett.123.250604} {\bibfield  {journal} {\bibinfo  {journal}
			{\prl}\ }\textbf {\bibinfo {volume} {123}},\ \bibinfo {pages} {250604}
		(\bibinfo {year} {2019})}\BibitemShut {NoStop}%
	\bibitem [{\citenamefont {Cirac}\ \emph {et~al.}(2020)\citenamefont {Cirac},
		\citenamefont {Perez-Garcia}, \citenamefont {Schuch},\ and\ \citenamefont
		{Verstraete}}]{Cirac2020}%
	\BibitemOpen
	\bibfield  {author} {\bibinfo {author} {\bibfnamefont {Ignacio}\ \bibnamefont
			{Cirac}}, \bibinfo {author} {\bibfnamefont {David}\ \bibnamefont
			{Perez-Garcia}}, \bibinfo {author} {\bibfnamefont {Norbert}\ \bibnamefont
			{Schuch}}, \ and\ \bibinfo {author} {\bibfnamefont {Frank}\ \bibnamefont
			{Verstraete}},\ }\bibfield  {title} {\enquote {\bibinfo {title} {{Matrix
					Product States and Projected Entangled Pair States: Concepts, Symmetries, and
					Theorems}},}\ }\href@noop {} {\  (\bibinfo {year} {2020})},\ \Eprint
	{http://arxiv.org/abs/2011.12127} {arXiv:2011.12127 [quant-ph]} \BibitemShut
	{NoStop}%
	\bibitem [{\citenamefont {Haegeman}\ and\ \citenamefont
		{Verstraete}(2017)}]{Haegeman2017}%
	\BibitemOpen
	\bibfield  {author} {\bibinfo {author} {\bibfnamefont {Jutho}\ \bibnamefont
			{Haegeman}}\ and\ \bibinfo {author} {\bibfnamefont {Frank}\ \bibnamefont
			{Verstraete}},\ }\bibfield  {title} {\enquote {\bibinfo {title}
			{Diagonalizing transfer matrices and matrix product operators: A medley of
				exact and computational methods},}\ }\href {\doibase
		10.1146/annurev-conmatphys-031016-025507} {\bibfield  {journal} {\bibinfo
			{journal} {Annual Review of Condensed Matter Physics}\ }\textbf {\bibinfo
			{volume} {8}},\ \bibinfo {pages} {355--406} (\bibinfo {year} {2017})},\
	\Eprint
	{http://arxiv.org/abs/https://doi.org/10.1146/annurev-conmatphys-031016-025507}
	{https://doi.org/10.1146/annurev-conmatphys-031016-025507} \BibitemShut
	{NoStop}%
	\bibitem [{\citenamefont {Fishman}\ \emph {et~al.}(2018)\citenamefont
		{Fishman}, \citenamefont {Vanderstraeten}, \citenamefont {Zauner-Stauber},
		\citenamefont {Haegeman},\ and\ \citenamefont {Verstraete}}]{fishman2018}%
	\BibitemOpen
	\bibfield  {author} {\bibinfo {author} {\bibfnamefont {M.~T.}\ \bibnamefont
			{Fishman}}, \bibinfo {author} {\bibfnamefont {L.}~\bibnamefont
			{Vanderstraeten}}, \bibinfo {author} {\bibfnamefont {V.}~\bibnamefont
			{Zauner-Stauber}}, \bibinfo {author} {\bibfnamefont {J.}~\bibnamefont
			{Haegeman}}, \ and\ \bibinfo {author} {\bibfnamefont {F.}~\bibnamefont
			{Verstraete}},\ }\bibfield  {title} {\enquote {\bibinfo {title} {Faster
				methods for contracting infinite two-dimensional tensor networks},}\ }\href
	{\doibase 10.1103/PhysRevB.98.235148} {\bibfield  {journal} {\bibinfo
			{journal} {Phys. Rev. B}\ }\textbf {\bibinfo {volume} {98}},\ \bibinfo
		{pages} {235148} (\bibinfo {year} {2018})}\BibitemShut {NoStop}%
	\bibitem [{\citenamefont {Nishino}\ \emph {et~al.}(1996)\citenamefont
		{Nishino}, \citenamefont {Okunishi},\ and\ \citenamefont
		{Kikuchi}}]{Nishino1996b}%
	\BibitemOpen
	\bibfield  {author} {\bibinfo {author} {\bibfnamefont {T.}~\bibnamefont
			{Nishino}}, \bibinfo {author} {\bibfnamefont {K.}~\bibnamefont {Okunishi}}, \
		and\ \bibinfo {author} {\bibfnamefont {M.}~\bibnamefont {Kikuchi}},\
	}\bibfield  {title} {\enquote {\bibinfo {title} {Numerical renormalization
				group at criticality},}\ }\href {\doibase 10.1016/0375-9601(96)00128-4}
	{\bibfield  {journal} {\bibinfo  {journal} {Physics Letters A}\ }\textbf
		{\bibinfo {volume} {213}},\ \bibinfo {pages} {69--72} (\bibinfo {year}
		{1996})}\BibitemShut {NoStop}%
	\bibitem [{\citenamefont {{Pollmann}}\ \emph {et~al.}(2009)\citenamefont
		{{Pollmann}}, \citenamefont {{Mukerjee}}, \citenamefont {{Turner}},\ and\
		\citenamefont {{Moore}}}]{Pollmann2009}%
	\BibitemOpen
	\bibfield  {author} {\bibinfo {author} {\bibfnamefont {F.}~\bibnamefont
			{{Pollmann}}}, \bibinfo {author} {\bibfnamefont {S.}~\bibnamefont
			{{Mukerjee}}}, \bibinfo {author} {\bibfnamefont {A.~M.}\ \bibnamefont
			{{Turner}}}, \ and\ \bibinfo {author} {\bibfnamefont {J.~E.}\ \bibnamefont
			{{Moore}}},\ }\bibfield  {title} {\enquote {\bibinfo {title} {{Theory of
					Finite-Entanglement Scaling at One-Dimensional Quantum Critical Points}},}\
	}\href {\doibase 10.1103/PhysRevLett.102.255701} {\bibfield  {journal}
		{\bibinfo  {journal} {\prl}\ }\textbf {\bibinfo {volume} {102}},\ \bibinfo
		{pages} {255701} (\bibinfo {year} {2009})}\BibitemShut {NoStop}%
	\bibitem [{\citenamefont {{Tagliacozzo}}\ \emph {et~al.}(2008)\citenamefont
		{{Tagliacozzo}}, \citenamefont {{de Oliveira}}, \citenamefont {{Iblisdir}},\
		and\ \citenamefont {{Latorre}}}]{Tagliacozzo2008}%
	\BibitemOpen
	\bibfield  {author} {\bibinfo {author} {\bibfnamefont {L.}~\bibnamefont
			{{Tagliacozzo}}}, \bibinfo {author} {\bibfnamefont {T.~R.}\ \bibnamefont {{de
					Oliveira}}}, \bibinfo {author} {\bibfnamefont {S.}~\bibnamefont
			{{Iblisdir}}}, \ and\ \bibinfo {author} {\bibfnamefont {J.~I.}\ \bibnamefont
			{{Latorre}}},\ }\bibfield  {title} {\enquote {\bibinfo {title} {Scaling of
				entanglement support for matrix product states},}\ }\href {\doibase
		10.1103/PhysRevB.78.024410} {\bibfield  {journal} {\bibinfo  {journal}
			{\prb}\ }\textbf {\bibinfo {volume} {78}},\ \bibinfo {eid} {024410} (\bibinfo
		{year} {2008})}\BibitemShut {NoStop}%
	\bibitem [{\citenamefont {Pirvu}\ \emph {et~al.}(2012)\citenamefont {Pirvu},
		\citenamefont {Vidal}, \citenamefont {Verstraete},\ and\ \citenamefont
		{Tagliacozzo}}]{pirvu2012matrix}%
	\BibitemOpen
	\bibfield  {author} {\bibinfo {author} {\bibfnamefont {B.}~\bibnamefont
			{Pirvu}}, \bibinfo {author} {\bibfnamefont {G.}~\bibnamefont {Vidal}},
		\bibinfo {author} {\bibfnamefont {F.}~\bibnamefont {Verstraete}}, \ and\
		\bibinfo {author} {\bibfnamefont {L.}~\bibnamefont {Tagliacozzo}},\
	}\bibfield  {title} {\enquote {\bibinfo {title} {Matrix product states for
				critical spin chains: Finite-size versus finite-entanglement scaling},}\
	}\href {\doibase 10.1103/PhysRevB.86.075117} {\bibfield  {journal} {\bibinfo
			{journal} {Phys. Rev. B}\ }\textbf {\bibinfo {volume} {86}},\ \bibinfo
		{pages} {075117} (\bibinfo {year} {2012})}\BibitemShut {NoStop}%
	\bibitem [{\citenamefont {Milsted}\ \emph {et~al.}(2013)\citenamefont
		{Milsted}, \citenamefont {Haegeman},\ and\ \citenamefont
		{Osborne}}]{milsted2013}%
	\BibitemOpen
	\bibfield  {author} {\bibinfo {author} {\bibfnamefont {Ashley}\ \bibnamefont
			{Milsted}}, \bibinfo {author} {\bibfnamefont {Jutho}\ \bibnamefont
			{Haegeman}}, \ and\ \bibinfo {author} {\bibfnamefont {Tobias~J.}\
			\bibnamefont {Osborne}},\ }\bibfield  {title} {\enquote {\bibinfo {title}
			{Matrix product states and variational methods applied to critical quantum
				field theory},}\ }\href {\doibase 10.1103/PhysRevD.88.085030} {\bibfield
		{journal} {\bibinfo  {journal} {Phys. Rev. D}\ }\textbf {\bibinfo {volume}
			{88}},\ \bibinfo {pages} {085030} (\bibinfo {year} {2013})}\BibitemShut
	{NoStop}%
	\bibitem [{\citenamefont {{Nishino}}\ and\ \citenamefont
		{{Okunishi}}(1996)}]{Nishino1996}%
	\BibitemOpen
	\bibfield  {author} {\bibinfo {author} {\bibfnamefont {T.}~\bibnamefont
			{{Nishino}}}\ and\ \bibinfo {author} {\bibfnamefont {K.}~\bibnamefont
			{{Okunishi}}},\ }\bibfield  {title} {\enquote {\bibinfo {title} {{Corner
					Transfer Matrix Renormalization Group Method}},}\ }\href {\doibase
		10.1143/JPSJ.65.891} {\bibfield  {journal} {\bibinfo  {journal} {Journal of
				the Physical Society of Japan}\ }\textbf {\bibinfo {volume} {65}},\ \bibinfo
		{pages} {891} (\bibinfo {year} {1996})}\BibitemShut {NoStop}%
	\bibitem [{\citenamefont {Or\'us}\ and\ \citenamefont
		{Vidal}(2009)}]{Orus2009}%
	\BibitemOpen
	\bibfield  {author} {\bibinfo {author} {\bibfnamefont {R.}~\bibnamefont
			{Or\'us}}\ and\ \bibinfo {author} {\bibfnamefont {G.}~\bibnamefont {Vidal}},\
	}\bibfield  {title} {\enquote {\bibinfo {title} {Simulation of
				two-dimensional quantum systems on an infinite lattice revisited: Corner
				transfer matrix for tensor contraction},}\ }\href {\doibase
		10.1103/PhysRevB.80.094403} {\bibfield  {journal} {\bibinfo  {journal} {Phys.
				Rev. B}\ }\textbf {\bibinfo {volume} {80}},\ \bibinfo {pages} {094403}
		(\bibinfo {year} {2009})}\BibitemShut {NoStop}%
	\bibitem [{\citenamefont {{Corboz}}\ \emph {et~al.}(2011)\citenamefont
		{{Corboz}}, \citenamefont {{White}}, \citenamefont {{Vidal}},\ and\
		\citenamefont {{Troyer}}}]{Corboz2011}%
	\BibitemOpen
	\bibfield  {author} {\bibinfo {author} {\bibfnamefont {P.}~\bibnamefont
			{{Corboz}}}, \bibinfo {author} {\bibfnamefont {S.~R.}\ \bibnamefont
			{{White}}}, \bibinfo {author} {\bibfnamefont {G.}~\bibnamefont {{Vidal}}}, \
		and\ \bibinfo {author} {\bibfnamefont {M.}~\bibnamefont {{Troyer}}},\
	}\bibfield  {title} {\enquote {\bibinfo {title} {{Stripes in the
					two-dimensional t-J model with infinite projected entangled-pair states}},}\
	}\href {\doibase 10.1103/PhysRevB.84.041108} {\bibfield  {journal} {\bibinfo
			{journal} {\prb}\ }\textbf {\bibinfo {volume} {84}},\ \bibinfo {pages}
		{041108} (\bibinfo {year} {2011})}\BibitemShut {NoStop}%
	\bibitem [{\citenamefont {Chang}(1976)}]{Chang1976}%
	\BibitemOpen
	\bibfield  {author} {\bibinfo {author} {\bibfnamefont {Shau-Jin}\
			\bibnamefont {Chang}},\ }\bibfield  {title} {\enquote {\bibinfo {title}
			{Existence of a second-order phase transition in a two-dimensional
				${\ensuremath{\varphi}}^{4}$ field theory},}\ }\href {\doibase
		10.1103/PhysRevD.13.2778} {\bibfield  {journal} {\bibinfo  {journal} {Phys.
				Rev. D}\ }\textbf {\bibinfo {volume} {13}},\ \bibinfo {pages} {2778--2788}
		(\bibinfo {year} {1976})}\BibitemShut {NoStop}%
	\bibitem [{\citenamefont {Brydges}\ \emph {et~al.}(1983)\citenamefont
		{Brydges}, \citenamefont {Frohlich},\ and\ \citenamefont
		{Sokal}}]{Brydges1983}%
	\BibitemOpen
	\bibfield  {author} {\bibinfo {author} {\bibfnamefont {D.~C.}\ \bibnamefont
			{Brydges}}, \bibinfo {author} {\bibfnamefont {J.}~\bibnamefont {Frohlich}}, \
		and\ \bibinfo {author} {\bibfnamefont {A.~D.}\ \bibnamefont {Sokal}},\
	}\bibfield  {title} {\enquote {\bibinfo {title} {{A NEW PROOF OF THE
					EXISTENCE AND NONTRIVIALITY OF THE CONTINUUM PHI**4 in two-dimensions AND
					PHI**4 in three-dimensions QUANTUM FIELD THEORIES}},}\ }\href {\doibase
		10.1007/BF01211157} {\bibfield  {journal} {\bibinfo  {journal} {Commun. Math.
				Phys.}\ }\textbf {\bibinfo {volume} {91}},\ \bibinfo {pages} {141--186}
		(\bibinfo {year} {1983})}\BibitemShut {NoStop}%
	\bibitem [{\citenamefont {Delcamp}\ and\ \citenamefont
		{Tilloy}(2020)}]{delcamp2020}%
	\BibitemOpen
	\bibfield  {author} {\bibinfo {author} {\bibfnamefont {Clement}\ \bibnamefont
			{Delcamp}}\ and\ \bibinfo {author} {\bibfnamefont {Antoine}\ \bibnamefont
			{Tilloy}},\ }\bibfield  {title} {\enquote {\bibinfo {title} {Computing the
				renormalization group flow of two-dimensional ${\ensuremath{\phi}}^{4}$
				theory with tensor networks},}\ }\href {\doibase
		10.1103/PhysRevResearch.2.033278} {\bibfield  {journal} {\bibinfo  {journal}
			{Phys. Rev. Research}\ }\textbf {\bibinfo {volume} {2}},\ \bibinfo {pages}
		{033278} (\bibinfo {year} {2020})}\BibitemShut {NoStop}%
	\bibitem [{\citenamefont {Kadoh}\ \emph {et~al.}(2019)\citenamefont {Kadoh},
		\citenamefont {Kuramashi}, \citenamefont {Nakamura}, \citenamefont {Sakai},
		\citenamefont {Takeda},\ and\ \citenamefont {Yoshimura}}]{Kadoh2018}%
	\BibitemOpen
	\bibfield  {author} {\bibinfo {author} {\bibfnamefont {Daisuke}\ \bibnamefont
			{Kadoh}}, \bibinfo {author} {\bibfnamefont {Yoshinobu}\ \bibnamefont
			{Kuramashi}}, \bibinfo {author} {\bibfnamefont {Yoshifumi}\ \bibnamefont
			{Nakamura}}, \bibinfo {author} {\bibfnamefont {Ryo}\ \bibnamefont {Sakai}},
		\bibinfo {author} {\bibfnamefont {Shinji}\ \bibnamefont {Takeda}}, \ and\
		\bibinfo {author} {\bibfnamefont {Yusuke}\ \bibnamefont {Yoshimura}},\
	}\bibfield  {title} {\enquote {\bibinfo {title} {{Tensor network analysis of
					critical coupling in two dimensional $\phi^{4}$ theory}},}\ }\href {\doibase
		10.1007/JHEP05(2019)184} {\bibfield  {journal} {\bibinfo  {journal} {JHEP}\
		}\textbf {\bibinfo {volume} {05}},\ \bibinfo {pages} {184} (\bibinfo {year}
		{2019})},\ \Eprint {http://arxiv.org/abs/1811.12376} {arXiv:1811.12376
		[hep-lat]} \BibitemShut {NoStop}%
	\bibitem [{Note1()}]{Note1}%
	\BibitemOpen
	\bibinfo {note} {Here we use the term 'observables' for quantities exhibiting
		a well posed continuum limit. The entanglement entropy is not an observable
		in the quantum information theoretic sense \cite
		{Nielsen-Chuang}}\BibitemShut {NoStop}%
	\bibitem [{\citenamefont {Calabrese}\ and\ \citenamefont
		{Cardy}(2004)}]{Calabrese_2004}%
	\BibitemOpen
	\bibfield  {author} {\bibinfo {author} {\bibfnamefont {Pasquale}\
			\bibnamefont {Calabrese}}\ and\ \bibinfo {author} {\bibfnamefont {John}\
			\bibnamefont {Cardy}},\ }\bibfield  {title} {\enquote {\bibinfo {title}
			{Entanglement entropy and quantum field theory},}\ }\href {\doibase
		10.1088/1742-5468/2004/06/p06002} {\bibfield  {journal} {\bibinfo  {journal}
			{Journal of Statistical Mechanics: Theory and Experiment}\ }\textbf {\bibinfo
			{volume} {2004}},\ \bibinfo {pages} {P06002} (\bibinfo {year}
		{2004})}\BibitemShut {NoStop}%
	\bibitem [{\citenamefont {Elias-Miro}\ \emph {et~al.}(2017)\citenamefont
		{Elias-Miro}, \citenamefont {Rychkov},\ and\ \citenamefont
		{Vitale}}]{Elias-Miro:2017}%
	\BibitemOpen
	\bibfield  {author} {\bibinfo {author} {\bibfnamefont {Joan}\ \bibnamefont
			{Elias-Miro}}, \bibinfo {author} {\bibfnamefont {Slava}\ \bibnamefont
			{Rychkov}}, \ and\ \bibinfo {author} {\bibfnamefont {Lorenzo~G.}\
			\bibnamefont {Vitale}},\ }\bibfield  {title} {\enquote {\bibinfo {title}
			{{High-Precision Calculations in Strongly Coupled Quantum Field Theory with
					Next-to-Leading-Order Renormalized Hamiltonian Truncation}},}\ }\href
	{\doibase 10.1007/JHEP10(2017)213} {\bibfield  {journal} {\bibinfo  {journal}
			{JHEP}\ }\textbf {\bibinfo {volume} {10}},\ \bibinfo {pages} {213} (\bibinfo
		{year} {2017})},\ \Eprint {http://arxiv.org/abs/1706.06121} {arXiv:1706.06121
		[hep-th]} \BibitemShut {NoStop}%
	\bibitem [{\citenamefont {Serone}\ \emph {et~al.}(2018)\citenamefont {Serone},
		\citenamefont {Spada},\ and\ \citenamefont {Villadoro}}]{serone2018}%
	\BibitemOpen
	\bibfield  {author} {\bibinfo {author} {\bibfnamefont {Marco}\ \bibnamefont
			{Serone}}, \bibinfo {author} {\bibfnamefont {Gabriele}\ \bibnamefont
			{Spada}}, \ and\ \bibinfo {author} {\bibfnamefont {Giovanni}\ \bibnamefont
			{Villadoro}},\ }\bibfield  {title} {\enquote {\bibinfo {title} {λϕ4 theory
				— part i. the symmetric phase beyond nnnnnnnnlo},}\ }\href {\doibase
		10.1007/JHEP08(2018)148} {\bibfield  {journal} {\bibinfo  {journal} {Journal
				of High Energy Physics}\ }\textbf {\bibinfo {volume} {2018}} (\bibinfo {year}
		{2018}),\ 10.1007/JHEP08(2018)148}\BibitemShut {NoStop}%
	\bibitem [{\citenamefont {Nielsen}\ and\ \citenamefont
		{Chuang}(2011)}]{Nielsen-Chuang}%
	\BibitemOpen
	\bibfield  {author} {\bibinfo {author} {\bibfnamefont {Michael~A.}\
			\bibnamefont {Nielsen}}\ and\ \bibinfo {author} {\bibfnamefont {Isaac~L.}\
			\bibnamefont {Chuang}},\ }\href@noop {} {\emph {\bibinfo {title} {Quantum
				Computation and Quantum Information: 10th Anniversary Edition}}},\ \bibinfo
	{edition} {10th}\ ed.\ (\bibinfo  {publisher} {Cambridge University Press},\
	\bibinfo {address} {USA},\ \bibinfo {year} {2011})\BibitemShut {NoStop}%
	\bibitem [{\citenamefont {Peskin}\ and\ \citenamefont
		{Schroeder}(1995)}]{PeskinSchroeder}%
	\BibitemOpen
	\bibfield  {author} {\bibinfo {author} {\bibfnamefont {Michael~E.}\
			\bibnamefont {Peskin}}\ and\ \bibinfo {author} {\bibfnamefont {Daniel~V.}\
			\bibnamefont {Schroeder}},\ }\href@noop {} {\emph {\bibinfo {title} {{An
					Introduction to quantum field theory}}}}\ (\bibinfo  {publisher}
	{Addison-Wesley},\ \bibinfo {address} {Reading, USA},\ \bibinfo {year}
	{1995})\BibitemShut {NoStop}%
\end{thebibliography}
%
\newpage
\pagestyle{empty}
\newpage

\appendix

\section{MPO Construction}\label{apendix:mpo}
We can write regularized partition function formally as a square lattice tensor network with on the vertices a tensor $T$ representing the degree of freedom of that vertex, and an operator, $t$, on the bonds:
\begin{equation}
\begin{split}
T=&\int\text{d}\phi\ket{\phi}\ket{\phi}\bra{\phi}\bra{\phi}\\
t=&\int\text{d}\phi\,\text{d}\phi^\prime e^{\frac{-1}{2}(\phi-\phi^\prime)^2-\frac{\mu^2}{8}(\phi^2+\phi^{\prime 2})-\frac{\lambda}{16}(\phi^4+\phi^{\prime 4})}\ket{\phi}\bra{\phi^\prime}
\end{split}
\end{equation}
Where we understand $\ket{\phi}$ to be the 'position' eigenbasis satisfying $\braket{\phi,\phi^\prime}=\delta(\phi-\phi^\prime)$. The on-site potential terms were evenly split over the four $t$'s neighboring a site. We next redefine the scale of $\phi$ by a factor of 2 to simplify the prefactors. Disregarding changes in overall scale we thus get
\begin{equation}
\begin{split}
T=&\int\text{d}\phi\ket{\phi}\ket{\phi}\bra{\phi}\bra{\phi}\\
t=&\int\text{d}\phi\,\text{d}\phi^\prime e^{-(\phi-\phi^\prime)^2-\frac{\mu^2}{4}(\phi^2+\phi^{\prime 2})-\frac{\lambda}{4}(\phi^4+\phi^{\prime 4})}\ket{\phi}\bra{\phi^\prime} \,.
\end{split}
\label{tensorsinphibasis}
\end{equation}
So now the lattice regularized partition function is expressed as a tensor network, but the bonds are now integrals. To overcome this issue we might consider expressing the basis elements $\ket{\phi}$ in terms of a new, countable, basis. The Hermite basis for example. However due to the difficulty in performing the double integral involved in $\bra{\psi_n}t\ket{\psi_m}$ this is not a good option. In stead, we first truncated the range of $\phi$ to an interval $[-L,L]$ such that all values of the exponential in \ref{tensorsinphibasis} within this interval do not differ more in scale than can be resolved at double precision, and consider the Fourier basis. The condition on the range of $\phi$ can be expressed as follows:
\begin{equation}
\begin{split}
e^{\frac{\mu^4}{8\lambda}}e^{-\frac{\mu^2}{2}\phi^2-\frac{\lambda}{2}\phi^4} & \geqslant \epsilon\\
\implies -\frac{\mu^2}{2}\phi^2-\frac{\lambda}{2}\phi^4& \geqslant\ln{\epsilon}-\frac{\mu^4}{8\lambda}
\end{split}
\end{equation}
where $\epsilon$ is the machine's precision. The above condition readily gives a lower bound on $L$, and it can be verified a posteriori that indeed the resulting MPO varies not with an increase in $L$.\\
$t$ can be easily expressed in the Fourier basis by discretizing $\phi$ in to $N$ equidistant points and performing a discrete Fourier transform. To make this more precise: consider following sampled version of $t$, $\tilde{t}$
\begin{equation}
\tilde{t}=\sum_{i,j}(\delta\phi)^2 e^{-(\phi_i-\phi_j)^2-\frac{\mu^2}{4}(\phi_i^2+\phi_j^2)-\frac{\lambda}{4}(\phi_i^4+\phi_j^4)} \ket{\phi_i}\bra{\phi_j}
\end{equation}
where $\delta\phi$ is the discretization size, $\phi_i$ are the discretization points and $\ket{\phi_i}$ are the basis states satisfying $\braket{\phi_i,\phi_j}=\frac{\delta_{i,j}}{\delta\phi}$ analogous to the definition of $\ket{\phi}$. We switch to a normalized basis $\ket{i}=\sqrt{\delta\phi}\ket{\phi_i}$ to normalize the inner product:
\begin{equation}
\tilde{t}=\sum_{i,j}\delta\phi e^{-(\phi_i-\phi_j)^2-\frac{\mu^2}{4}(\phi_i^2+\phi_j^2)-\frac{\lambda}{4}(\phi_i^4+\phi_j^4)} \ket{i}\bra{j}
\end{equation}
Performing a discrete Fourier transformation (by means of an fft) an approximation of $t$ expressed in the Fourier basis is obtained. \\
After convergence in Fourier modes, it becomes clear $t$ is incredibly rank deficient, showing that the better basis would actually be the eigenbasis. \\
The eigenvalues of $t$ --which are also the eigenvalues of the Fourier transformed $t$-- are always real and positive, at least for the couplings $\mu^2$ and $\lambda$ we will consider. We can thus perform following decomposition:
\begin{equation}
\tilde{t}=U^\dagger \sqrt{S} \sqrt{S} U = M^\dagger M
\end{equation}
The left space of $M$ is the eigenbasis of $t$, the right index however is a discrete sampled basis so the eigenvectors can not really converge. We could fft the right basis to get a more 'proper' basis that will converge, but it will actually be more beneficial not to.\\
It does not matter whether the eigenbasis of $t$ has converged, it only matters that the MPO has converged. We thus define $\tilde{T}$ analogous to how we defined $\tilde{t}$:
\begin{equation}
\begin{split}
\tilde{T}&=\sum_i\delta\phi\ket{\phi_i}\ket{\phi_i}\bra{\phi_i}\bra{\phi_i}\\
&=\sum_i(\delta\phi)^{-1}\ket{i}\ket{i}\bra{i}\bra{i}
\end{split}
\end{equation}
so $\tilde{T}$ is something like a four index ghz state rescaled by a factor $(\delta\phi)^{-1}$. It's now easy to write our MPO:
\begin{equation}
MPO=\sum_i(\delta\phi)^{-1}M\ket{i}M\ket{i}\bra{i}M^\dagger\bra{i}M^\dagger
\end{equation}
This object has no exposed sample dimensions, in fact all four indices are expressed in the eigenbasis of $t$. We observe that a slightly larger sample size is required to converge than was needed to get the fft of $\tilde{t}$ to converge, some $N=300-400$ tends to be more than enough to reach machine precision.\\
Note that, due to the particularly simple form of $\tilde{T}$, this object can easily be made, even if $N$ is so large that $\tilde{T}$ can not be fit into memory. In the Fourier basis this would be much more challenging.\\
Local observables $\braket{f(\phi_j)}$ are now easily calculated by inserting following tensor in the tensor network on position $j$.
\begin{equation}
T^\prime_{f}=\sum_i(\delta\phi)^{-1} f(\sqrt{2}\phi_i)M\ket{i}M\ket{i}\bra{i}M^\dagger\bra{i}M^\dagger\,,
\end{equation}
where the factor $\sqrt{2}$ is due to the rescaling of $\phi$ at the start.
\\


\section{Fitting a Scaling Function} \label{apendix:scalingfunctionfit}
Using a data collapse to fit something is no easy task, it has been the subject of many papers in the past \onlinecite{1} and continues to be the subject of research today. We use a fairly rudimentary approach where we construct a reasonable ansatz for the scaling functions based on their known properties and then optimize the fit in a total least square sense.\\
Preliminary inspection of the collapse plots and experience tells us we should seek to parameterize a set of functions with following properties:
\begin{enumerate}
	\item A non-analyticity at a point $x_0$.
	\item Goes to power law behavior far from zero.
	\item Some freedom in the shape around $x_0$.
	\item Differentiable everywhere but $x_0$.
\end{enumerate}
The form we chose that fits all these criteria is the following:
\begin{equation}
	f(x) = \begin{cases} 
		A(x-x_0)^{\beta +\sum_{i=1}^{N} \frac{a_i}{(x-x_0-|b_i|)^{c_i}+d_i}} & \text{if $x \leq x_0$} \\  
		A^\prime(x-x_0)^{\beta +\sum_{i=1}^{N} \frac{a^\prime_i}{(x-x_0+|b_i^\prime|)^{c^\prime_i}+d^\prime_i}} & \text{if $x > x_0$}
	\end{cases} 
\end{equation}
Scaling functions have no preferred axes, in the sense that collapsed points can have the wrong $x$-value or $y$-value. Additionally scaling functions can have a widely varying gradient, so the usual cost function of $y$-direction distance will tend to give a higher weight to regions of high gradient than to regions of low gradient, for which there is no justification.\\
We will therefore take the orthogonal distance between the points and the fit function as error measure, which also best coincides with the intuitive visual notion of collapsing. This is the main motivation for the criterion of differentiability we imposed on the scaling function ansatz, knowledge of the derivative of the scaling function enables accurate and efficient calculation of the orthogonal distance from a point to the function. 
\par
We thus obtain a cost function that we optimize with a standard tool based on gradient descent. The total least squares distance and the differentiable scaling function ansatz both aid significantly in smoothing out the cost function, but still there remain local minima, discontinuities and plateaus of vanishing gradient. It is therefore necessary to restart the optimization many times over, each time with slight random perturbations, to be sure of convergence to the true minimum.

\section{Renormalization of $\phi^3$}\label{appendix:phi3}

\begin{figure}
	\centering
	\includegraphics[width=0.8\linewidth]{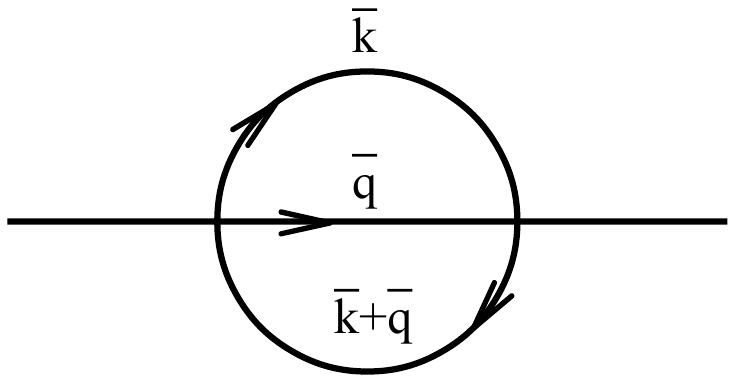}
	\caption{The setting sun diagram}
	\label{fig:settingsun}
\end{figure}

As the tadpole is the only UV-divergence, it follows immediately that the sole UV-divergent contribution to $\phi^3$, comes from the tadpole in $\braket{\phi^2}$ for the disconnected part $3\braket{\phi}\braket{\phi^2}$. As this divergence is universal, we can evaluate it e.g. by simply considering a cut-off $\Lambda$ in momentum space\cite{PeskinSchroeder}: \be \braket{\phi^2}=\int\! \frac{d^2k}{(2\pi)^2} \frac{1}{k^2+\mu_R^2}\approx \frac{1}{4\pi}\log\Lambda^2=-\frac{1}{4\pi}\log\lambda \,,\ee
where the last equation holds for the lattice regularization (\ref{eq:partitionfunction}), with $\lambda=a^2=1/\Lambda^2$. This then shows that the regularized $\tilde{\phi}^3=\phi^3+\frac{3}{4\pi}\log(\lambda)\phi\,$ is indeed UV-finite.    

\section{Corrections to UV-scaling from the setting sun diagram}\label{appendix:settingsun}
Upon rewriting $\alpha=\mu_R^2/\lambda$, it is apparent that the corrections in (\ref{eq:alphacorr}) are in fact corrections to the mass renormalization $\delta \mu^2\equiv \mu^2-\mu_R^2$. At one loop this counterterm follows from the tapdole diagram, giving rise to a divergent counterterm $\delta\mu^2 \propto \lambda \log (\Lambda/\mu_R^2)$, where $\Lambda$ is the UV-cut off ($=1/a$ for our lattice regularization) and where we work in physical (regularization independent units), with e.g. a mass dimension 2 for $\lambda$ and $\mu^2$. This is the only UV-divergence, all higher order diagrams are finite. Still, for finite $\Lambda$, also these higher order diagrams will of course have a $\Lambda$-dependence. Here we will isolate the $\Lambda$-dependent contribution that arises from the two loop setting-sun diagram (Fig. \ref{fig:settingsun}).

First of all, from a dimensional analysis we can infer the following general behaviour for the contribution to the zero-momentum inverse propagator $M(0)$ coming from this diagram: 

\be M_2(0)= \frac{\lambda^2}{\mu_R^2}\times(c_0+ c_1 \frac{\mu_R^2}{\Lambda^2}+c_2 (\frac{\mu_R^2}{\Lambda^2})^2+\ldots )\,. \label{eq:setting-sunbehaviour} \ee

Here, $c_0$ is the finite term that remains in the $\Lambda\rightarrow \infty$ limit, and $c_1$ captures the dominant $\Lambda$-dependence. But notice that $c_1$ itself can in principle have a logarithmic divergence $c_1\sim (\log\frac {\Lambda^2}{\mu_R^2})^{n}$ for some power $n$, since the $\frac{1}{\Lambda^2}$ pre-factor will still lead to a vanishing total term in the $\Lambda \rightarrow \infty$ limit. As we will show for our lattice regularization, this is precisely the case, with $n=1$. This then leads to an 'RG-improved' correction to the counterterm, compensating for the $\Lambda$-dependence of $M_2(0)$: \be \delta\mu^2_2= -\frac{\lambda^2}{\Lambda^2}\times c_1\,\,\propto \frac{\lambda^2}{\Lambda^2}\log(\Lambda^2/\mu_R^2)\,.\ee Restoring the lattice units (but keeping the same notation for $\lambda,\mu^2$), this becomes:
\be \delta \alpha = \frac{\delta\mu_2^2}{\lambda}\sim \lambda \log \lambda\,, \ee
which is precisely the leading correction in eq. (\ref{eq:alphacorr}).

Let us now show that $M_2(0)$ has indeed the described behaviour. For our lattice regularization (\ref{eq:partitionfunction} the setting-sun diagram contribution reads:
\be M_2(0)= \lambda^2 \prod_i\int_\pi^\pi\! dk_i \prod_i \int_\pi^\pi\! dq_i\left(P(k_i) P(q_i) P(k_i+q_i)\right)\, \label{eq:settingsun}\ee
with the propagators \be P(k_i)=\frac{1}{4 \sin^2(k_1/2)+4\sin^2(k_2/2)+\mu_R^2}\,.\ee Upon inspection of (\ref{eq:setting-sunbehaviour}), it is clear that the $c_0$ and $c_1$ terms can be identified from the singular $\mu_R^2\rightarrow 0$ behaviour of the setting-sun. This singular behavior emerges from the integration regions $k_i,q_i,k_i+q_i\approx 0$. To isolate these dominant $\mu_R^2\rightarrow 0$ contributions, we have expanded each of the propagators as: \be P(k_i)\approx \frac{1}{k_1^2+k_2^2+\mu_R^2}+\frac{1}{12}\frac{k_1^4+k_2^4}{(k_1^2+k_2^2+\mu_R^2)^2}+\ldots \,.\ee
It is then fairly easy to show that the contribution to the setting-sun from the lowest order terms for each of the propagators results in a term $\propto \mu_R^{-2}$, while taking the next term for one of the propagators gives a $\propto \log(\mu_R^2)$ contribution. By numerically evaluating the corresponding integrals we find the following dominant ($\mu_R^2\rightarrow 0$) setting-sun contribution:
\be M_2(0)\approx\lambda^2\left(\frac{23.1}{\mu_R^2}-16.8\log(\mu_R^2)\right),\ee
This behaviour, including the numerical factors, was confirmed by a fit to the numerically evaluated full exact expression (\ref{eq:settingsun}).
Notice, that while this setting-sun evaluation indeed allows us to infer the presence of a $\lambda\log\lambda$ term in (\ref{eq:alphacorr}), it does not predict a value of the pre-factor. Indeed, again from a dimensional analysis, one can write down generic higher order corrections that will also lead to an effective $\lambda\log \lambda$ correction in (\ref{eq:alphacorr}).


\end{document}